\documentclass{elsart3p}

\usepackage{graphicx,natbib} 
\usepackage{amsmath,amssymb,float} 
\usepackage{epstopdf} 

\begin{document}

\def\K{{\bf{K}}} 
\def\Q{{\bf{Q}}} 
\def\q{{\bf{q}}} 
\def\Gbar{\bar{G}} 
\def\tk{\tilde{\bf{k}}} 
\def\k{{\bf{k}}} 
\def\kt{{\tilde{\bf{k}}}} 
\def\p{{\bf{p}}} 
\def\pp{{\bf{p}}^\prime} 
\def\Gpp{\Gamma^{pp}} 
\def\Gph{\Gamma^{ph}} 
\def\Phid{\Phi_d(\k,\omega_n)} 
\def\ld{\lambda_d(T)} 
\def\n{\langle n \rangle } 
\def\dw{d_{x^2-y^2}} 
\begin{frontmatter}
	
	
	\title{Understanding High-Temperature Superconductors with Quantum Cluster Theories}
	
	
	\author[1]{T.A.~Maier} 
	\author[2]{M.S.~Jarrell} 
	\author[3]{D.J.~Scalapino} \address[1]{Computer Science and Mathematics Division, Oak Ridge National Laboratory, Oak Ridge, TN 37831, USA} \address[2]{Department of Physics, University of Cincinnati, Cincinnati, OH 45221, USA} \address[3]{University of California, Santa Barbara, CA 93106, USA} 
	\begin{abstract}
		
		Quantum cluster theories are a set of approaches for the theory of correlated and disordered lattice systems, which treat correlations within the cluster explicitly, and correlations at longer length scales either perturbatively or within a mean-field approximation. These methods become exact when the cluster size diverges, and most recover the corresponding (dynamical) mean-field approximation when the cluster size becomes one. Here we will review systematic dynamical cluster simulations of the two-dimensional Hubbard model, that display phenomena remarkably similar to those found in the cuprates, including antiferromagnetism, superconductivity and pseudogap behavior. We will then discuss results for the structure of the pairing mechanism in this model, obtained from a combination of dynamical cluster results and diagrammatic techniques. 
	\end{abstract}
\end{frontmatter}

\section{Introduction} \label{sec:Intro}

Despite two decades of intense studies, high-temperature superconductivity remains a mystery and represents one of the most important outstanding problems in condensed matter science today. The two-dimensional Hubbard model \cite{anderson:rvb,zhang:88} of the cuprates and related models have been studied extensively with numerical techniques, such as exact diagonalization and quantum Monte Carlo (QMC), on finite size lattices \cite{dagotto:rmp}. These studies indicate that d-wave pairing correlations develop as the temperature is lowered (see e.g. Ref.~\cite{bulut:93}). However, the numerical expense or minus sign problem have limited finite size techniques to lattices too small and temperatures too high to draw conclusive evidence for the existence of d-wave superconductivity in these models \cite{scalapino:99}.

Recently, quantum cluster theories \cite{maier:rev} have been used successfully to shed new light on the physics of the 2D Hubbard model. In these approaches a finite size cluster is embedded in a self-consistent host, designed to represent the rest of the system. The Cellular Dynamical Mean Field Theory \cite{kotliar:cdmft} uses a cluster in real space with open boundary conditions, while the Dynamical Cluster Approximation (DCA) \cite{hettler:dca1} treats a cluster in reciprocal space. Correlations within the cluster are treated explicitly, while longer-ranged correlations beyond the cluster size are described in a mean-field. In contrast to finite size techniques, these approaches remain a non-trivial and physically relevant approximation to the thermodynamic limit even for small cluster sizes. For a cluster consisting of only a single site, these techniques reduce to the dynamical mean field theory \cite{pruschke:dmftrev} which has been used successfully to investigate the Mott metal to insulator transition. To describe d-wave pairing, however, one needs to use a cluster of at least four sites, which represents the mean-field solution for d-wave superconducting order since it does not allow for pairfield phase fluctuations \cite{maier:schm}. 
\begin{figure}
	[htpb] 
	\begin{center}
		\includegraphics[width=3in]{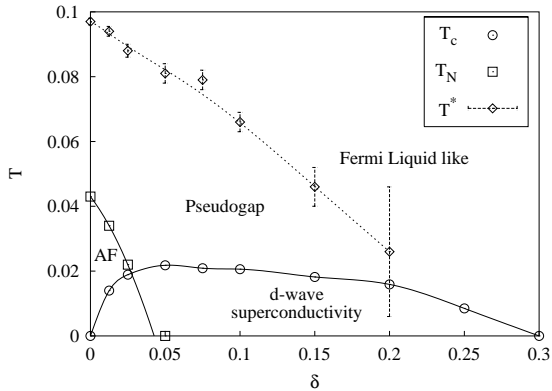} 
	\end{center}
	\caption{The temperature-doping phase diagram of the 2D Hubbard model obtained with a 4-site DCA calculation. (Jarrell {\it et al.} \cite{jarrell:dca2})} \label{fig:DCApd} 
\end{figure}

\begin{figure}
	[htb] \centerline{ 
	\includegraphics*[width=2.7in]{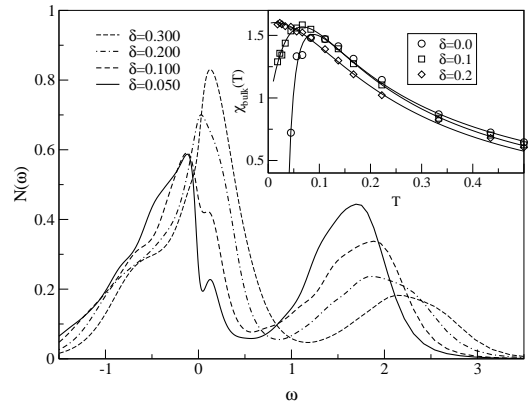}} \caption{DOS for various dopings $\delta=1-\n$ in the 2D Hubbard model at $T=0.092t$ and $U=8t$ calculated with DCA/QMC for a 4-site cluster, $N_c=4$. Inset: Uniform spin-susceptibility as a function of temperature. Energies are in units of $4t$. (Jarrell {\it et al.} \cite{jarrell:dca2})} \label{fig:res2dhmpg2} 
\end{figure}

To illustrate the power of quantum cluster theories for even small cluster sizes, we review in Fig.~\ref{fig:DCApd} the temperature-doping phase diagram of a 2D Hubbard model with nearest neighbor hopping $t$ and Coulomb repulsion $U=8t$ obtained with a 4-site DCA calculation \cite{jarrell:dca2}. The phase diagram is remarkably similar to the universal cuprate phase diagram. It displays an antiferromagnetic phase near half-filling, a d-wave superconducting phase at finite doping and, as shown in Fig.~\ref{fig:res2dhmpg2}, pseudogap behavior in the density of states and spin susceptibiliy at higher temperatures. Very similar results obtained with CDMFT \cite{kyung:prb06} and cluster perturbation theory (a quantum cluster theory where the host is not treated self-consistently) \cite{senechal:cluster} have confirmed this picture. 

In this paper, we will review DCA simulations for a 2D Hubbard model with nearest-neighbor hopping $t$ and an on-site Coulomb interaction $U$. We will review evidence that superconductivity persists in larger clusters \cite{maier:schm}, as well as recent studies of the nature of the pairing interaction mediating d-wave superconductivity \cite{maier:pairmech,maier:pairint} in this model. In the following section~\ref{sec:QCT} we review quantum cluster theories with a focus on the dynamical cluster approximation. In Sec.~\ref{sec:SC} we explore the cluster size and shape dependence of the pairing correlations and the superconducting transition temperature $T_c$. Then, in Sec.~\ref{sec:PM} we investigate the nature of the pairing interaction by examining the leading eigenvalue of the homogeneous particle-particle Bethe-Salpeter equation and the momentum and frequency dependence of the corresponding eigenfunction. Sec.~\ref{sec:Conc} contains our conclusions.

\section{Quantum Cluster Theories: Dynamical Cluster Approximation} \label{sec:QCT}

Quantum cluster approaches share the general idea to approximate the effects of correlations in the bulk lattice with those on a finite size quantum cluster \cite{maier:rev}. This enables a mapping of the bulk lattice problem to an effective cluster embedded in a self-consistent bath designed to represent the remaining degrees of freedom. In contrast to finite system simulations which determine the exact state of a finite system, quantum cluster techniques give approximate results to the bulk thermodynamic result. In this approximation, short-ranged correlations within the cluster are treated explicitly, while the longer-ranged physics is described on the mean-field level. By increasing the cluster size, the quantum cluster theories systematically interpolate between the single-site dynamical mean-field result and the exact result, while remaining an approximation to the thermodynamic limit for finite cluster size. In the following we will focus on one of the quantum cluster methods, the DCA, which has been used to obtain the results presented in this manuscript. 
\begin{figure}
	[htpb] 
	\begin{center}
		\includegraphics[width=1.75in]{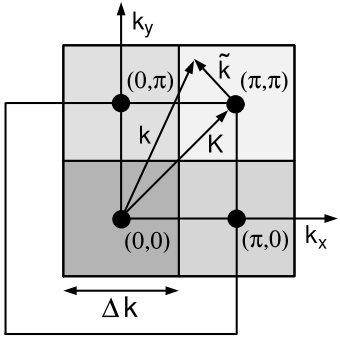} 
	\end{center}
	\caption{In the DCA the Brillouin zone is divided into $N_c$ cells each represented by a cluster momentum $\K$. Irreducible quantities such as the single-particle self-energy $\Sigma$ and two-particle irreducible vertices are constructed from coarse-grained propagators $\Gbar(\K)$ that are averaged over the momenta $\tk$ within the cell represented by $\K$.} \label{fig:DCABZ} 
\end{figure}

The essential assumption of the DCA \cite{hettler:dca1} is that short-range quantities, such as the single-particle self-energy $\Sigma$, and its functional derivatives, the two-particle irreducible vertex functions, are well represented as diagrams constructed from a coarse-grained propagator $\Gbar$. To define $\Gbar$, the Brillouin zone in two dimensions is divided into $N_c=L^2$ cells of size $2\pi/L^2$. As illustrated in Fig.~\ref{fig:DCABZ}, each cell is represented by the cluster momentum $\K$ in its center. The coarse-grained Green function $\Gbar(\K)$ is then obtained from an average over the $N/N_c$ wave-vectors $\kt$ within the cell surrounding $\K$, 
\begin{eqnarray}
	\label{eq:Gbar} \Gbar(\K,\omega_n) = && \nonumber\\
	&&\hspace*{-1cm}\frac{N_c}{N}\sum_{\kt} \frac{1}{i\omega_n-\epsilon_{\K+\kt} +\mu-\Sigma_c(\K,\omega_n)}. 
\end{eqnarray}

\noindent Here the self-energy for the bulk lattice $\Sigma(\K+\kt,\omega_n)$ has been approximated by the cluster self-energy $\Sigma_c(\K,\omega_n)$. Consequently, the compact Feynman diagrams constructed from $\Gbar(\K,\omega_n)$ collapse onto those of an effective cluster problem embedded in a host which accounts for the fluctuations arising from the hybridization between the cluster and the rest of the system. The non-interacting part of the effective cluster action is then defined by the cluster-excluded inverse Green's function 
\begin{eqnarray}
	\label{eq:Gscript} {\mathcal G}^{-1}(\K,\omega_n) = \Gbar^{-1}(\K,\omega_n)+\Sigma_c(\K,\omega_n) 
\end{eqnarray}

\noindent which accounts for the hybridization between the cluster and the host. Given ${\mathcal G}^{-1}(\K,\omega_n)$ and the interaction on the cluster $U\sum_i n_{i\uparrow}n_{i\downarrow}$, one can then set up a Hirsch-Fye quantum Monte Carlo algorithm \cite{hirsch:QMC} to calculate the cluster Green's function $G_c(\K,\omega_n)$ and from it the cluster self-energy $\Sigma_c(\K,\omega_n)$ which is used in Eq.~(\ref{eq:Gbar}) to re-calculate the coarse-grained Green function $\Gbar(\K,\omega_n)$ \cite{jarrell:dca3,maier:rev}. This process is then iterated to convergence.

Since a determinantal Monte Carlo method is used to solve the effective cluster problem, there is also a sign problem for the doped Hubbard model. However, the coupling of the cluster to the self-consistent host significantly reduce the sign problem so that lower temperatures can be reached \cite{jarrell:dca3}.

With the Hirsch-Fye QMC algortihm, one may also calculate the cluster two-particle Green's function $G_{c2}(K_4, K_3; K_2, K_1)$ with $K=({\K}, i\omega_n, \sigma)$. Using $G_c(K)$ and $G_{c2}(K_4, K_3; K_2, K_1)$, one can extract the cluster four-point vertex $\Gamma$ from 
\begin{eqnarray}
	\label{six} G_{c2} (K_4, K_3; K_2, K_1) = \hspace*{-3cm} &&\nonumber\\
	&-& G_c(K_1)\, G_c(K_2)\, \left[\delta_{K_1, K_4} \delta_{K_2, K_3} - \delta_{K_1, K_3} \delta_{K_2, K_4}\right]\nonumber \\
	&+& \frac{T}{N}\ \delta_{K_1+K_2, K_3+K_4} G_c(K_4)\, G_c(K_3) \nonumber \\
	&\times& \Gamma\, (K_4,K_3; K_2, K_1)\,G_c(K_2)\, G_c(K_1)\, . 
\end{eqnarray}

Then, using $G_c$ and $\Gamma$, one can determine the corresponding irreducible particle-hole and particle-particle vertices on the cluster, $\Gamma^{ph}$ and $\Gamma^{pp}$, respectively, from the corresponding Bethe-Salpeter equations. Two-particle Green's functions for the original bulk lattice, $G_2(k_4,k_3,k_2,k_1)$, with $k=(\k,i\omega_n,\sigma)$, are then obtained by using the cluster irreducible vertices, $\Gamma^{ph}$ and $\Gamma^{pp}$, as approximations for the lattice irreducible vertices. Lattice susceptibilities such as the pair-field susceptibility discussed in Sec.~\ref{sec:SC} may then be obtained from $G_2(k_4,k_3,k_2,k_1)$. n

\section{Superconductivity in the 2D Hubbard model} \label{sec:SC}

With increasing cluster size, quantum cluster theories including the DCA progressively include longer-ranged fluctuations while retaining some mean-field character \cite{maier:rev}. With respect to the 4-site phase diagram shown in Fig.~\ref{fig:DCApd}, larger clusters are thus expected to systematically drive the Ne{\'e}l temperature to zero and hence recover the Mermin-Wagner theorem. In contrast, superconductivity may persist as Kosterlitz-Thouless order, even in the large cluster limit \cite{maier:schm}. 

A measure of the strength of the d-wave pairing correlations is given by the d-wave pairfield susceptiblity 
\begin{eqnarray}
	P_d=\int_0^\beta d\tau \langle \Delta_d(\tau) \Delta_d^\dagger(0) \rangle \label{eq:Pd} 
\end{eqnarray}

\noindent with $\Delta_d^\dagger = 1/2\sqrt{N}\sum_{l,\delta} (-1)^\delta c^\dagger_{l\uparrow} c^\dagger_{l+\delta\downarrow}$. Here, $\delta$ sums over the four near-neighbor sites of $l$. In the DCA, $P_d$ is calculated from the two-particle lattice Green's function $G_2(k_4,k_4,k_2,k_1)$ discussed in Sec.~\ref{sec:QCT}. The divergence of $P_d$ indicates a phase transition to a d-wave superconducting state. 
\begin{figure}
	[htpb] 
	\begin{center}
		\includegraphics[width=2.5in]{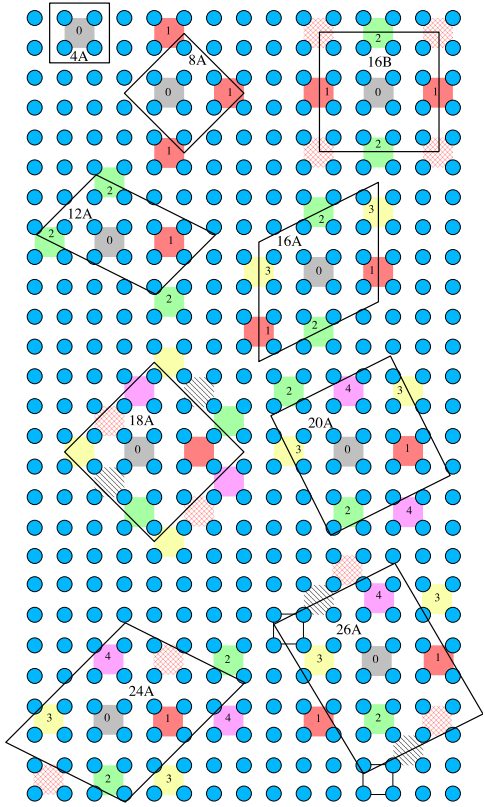} 
	\end{center}
	\caption{Different cluster sizes and shapes used in the DCA study of pairing in the 2D Hubbard model. (Maier {\it et al.} \cite{maier:schm})} \label{fig:clusters} 
\end{figure}

To select different cluster sizes and shapes, it is useful to adopt the cluster selection criteria originally introduced by Betts {\it et al.} \cite{dbetts:2d} in a numerical study of the 2D Heisenberg model. For the Heisenberg model, Betts {\em et al.} developed a grading scheme to determine which clusters should be used. An important selection criterium was the completeness of near-neighbor shells compared to the infinite lattice. They found that a finite size scaling analysis was greatly improved when only the most perfect clusters were used. In Ref.~\cite{maier:schm}, some of us generalized Bett's arguments to generate clusters appropriate to study superconductivity. For a d-wave order parameter, one needs to take into account its non-local 4-site plaquette structure. 
\begin{table}
	[h] \caption{Number of independent neighboring d-wave plaquettes $Z_d$ and the values of $T_c^{\rm lin}$ obtained from linear fits of the pair-field susceptibility in Fig.~\ref{fig:SC}. (Maier {\it et al.} \cite{maier:schm})} 
	\begin{center}
		\begin{tabular}
			{ccl} \hline\hline Cluster & $Z_d$ & $T_c^{\rm lin}/t$ \\
			\hline 4 & 0 (MF) & 0.056\\
			8A & 1 & -0.006 \\
			18A & 1 & -0.022 \\
			12A & 2 & 0.016 \\
			16B & 2 & 0.015 \\
			16A & 3 & 0.025$\pm$0.002 \\
			20A & 4 & 0.022 \\
			24A & 4 & 0.020 \\
			26A & 4 & 0.023 \\
			\hline\hline 
		\end{tabular}
	\end{center}
	\label{tab:1} 
\end{table}

Fig.~\ref{fig:clusters} shows the arrangement of independent d-wave plaquettes in the clusters. Denoting the number of independent near-neighbor plaquettes by $Z_d$, the inifinite lattice has $Z_d=4$. The 4-site cluster encloses just one d-wave plaquette, so $Z_d=0$. In this case, the effective action does not contain any pairfield fluctuations. Thus, the 4-site cluster represents the mean-field result for d-wave pairing and hence overstimates $T_c$. In the 8A cluster, there is room for one more d-wave pair ($Z_d=1$). The same neighboring plaquette is adjacent to its partner on all four sides, phase fluctuations are replicated and hence overestimated, and thus $T_c$ is suppressed. For the 16B cluster, $Z_d=2$, while for the oblique 16A cluster, $Z_d=3$. Thus, one expects the pairing correlations in the 16B cluster to be suppressed relative to those in the 16A cluster. The number $Z_d$ of independent neighboring d-wave plaquettes for the clusters in Fig.~\ref{fig:clusters} are listed in Table~\ref{tab:1}. The 20A, 24A and 26A clusters all have $Z_d=4$ and thus are expected to show the most accurate results. 
\begin{figure}
	[h] 
	\begin{center}
		\includegraphics[height=3.2in,angle=-90]{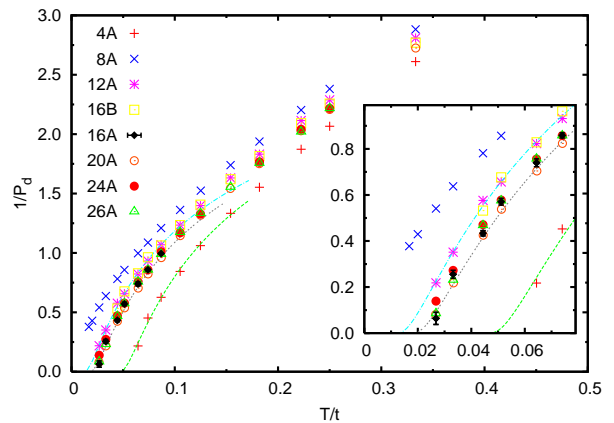} 
	\end{center}
	\caption{The inverse pairfield susceptibility versus temperature of the 2D Hubbard model calculated with DCA for different cluster sizes and shapes. (Maier {\it et al.} \cite{maier:schm})} \label{fig:SC} 
\end{figure}

Fig.~\ref{fig:SC} shows results for the inverse pairfield susceptibility $1/P_d$ versus temperature for $U=4t$ and $\n=0.9$. As expected the mean-field $N_c=4$ result overestimates $T_c$, and the results for the 8A and 18A clusters, both of which have $Z_d=1$, do not give a positive $T_c$. However, results on larger clusters with $Z_d\ge 3$ fall nearly on the same curve. The results suggest that the 2D Hubbard model with $U=4t$ and $\n=0.9$ has a transition to a d-wave superconducting phase at a finite temperature $T_c$. Due to the mean-field character of the DCA, one expect mean-field behavior close to $T_c$. The results for $T_c$ of a mean-field linear fit of $1/P_d$ close to the transition are listed in Table~\ref{tab:1}. For all cluster with $Z_d\ge 3$, a transition temperature $T_c\approx 0.023t\pm0.002 t$ is found.

\section{The Structure of the Effective Pairing Interaction} \label{sec:PM}

\begin{figure}
	[htpb] 
	\begin{center}
		\includegraphics[width=2.8in]{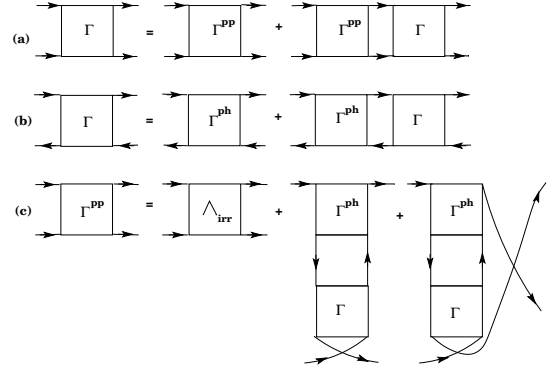} 
	\end{center}
	\caption{Bethe-Salpeter equations for (a) the particle-particle and (b) the particle-hole channels showing the relationship between the full vertex $\Gamma$, the particle-particle irreducible vertex $\Gpp$ and the particle-hole irreducible vertex $\Gph$, respectively. (c) Decomposition of the irreducible particle-particle vertex $\Gpp$ into a fully irreducible two-fermion vertex $\Lambda_{\rm irr}$ plus contributions from the particle-hole channels. All diagrams represent DCA cluster quantities, including the Green's function legs. (Maier {\it et al.} \cite{maier:pairmech})} \label{fig:bseqs} 
\end{figure}

As discussed in the previous section, DCA calculations find evidence for a finite temperature d-wave superconducting phase in the 2D Hubbard model. Here we review insight obtained by combining numerical results and diagrammatic methods to determine the structure of the interaction mediating the d-wave pairing \cite{maier:pairmech,maier:pairint}. The basic idea is to focus on the 4-point vertex $\Gamma$ shown in Fig.~\ref{fig:bseqs} calculated with a DCA/QMC simulation from Eq.~\ref{six}. From this vertex and results for the single-particle Green's function, one can determine the irreducible particle-particle, $\Gpp$, and particle-hole, $\Gph$, vertices using the Bethe-Salpeter equations shown in Fig.~\ref{fig:bseqs}a and b. Since the Monte Carlo results for $\Gamma$ satisfy crossing symmetry, the pairing interaction $\Gpp$ may be related to the particle-hole channels through the equation shown in Fig.~\ref{fig:bseqs}. 

In order to determine the structure of the dominant correlations in the system, the cluster irreducible vertices and the lattice single-particle Green's function are used to calculate the eigenvalues and eigenfunctions of the homogeneous Bethe-Salpeter equation. For example in the particle-particle channel 
\begin{eqnarray}
	-\frac{T}{N}\ \sum_{k^\prime} \Gamma^{\rm pp} (K|K') G_\uparrow (k^\prime)\, G_\downarrow (-k^\prime)\, \phi_\alpha (K^\prime) = && \nonumber\\
	&&\hspace{-1.9cm}\lambda_\alpha \phi_\alpha (K) \label{seven} 
\end{eqnarray}

\noindent with a similar equation using $\Gamma^{ph}$ for the particle-hole channel. Here, $\Gamma^{pp}(K|K') = \Gamma^{pp}(K,-K;K',-K')$ and the sum over $k^\prime$ denotes a sum over both momentum $\k^\prime=\K'+\tk'$ and Matsubara $\omega_{n^\prime}$ variables. By assumption, irreducible quantities like $\Gamma^{pp}$ and $\phi_\alpha$ do not depend on $\tk^\prime$, allowing to coarse-grain the Green function legs, yielding an equation that depends only on coarse-grained and cluster quantities 
\begin{eqnarray}
	-\frac{T}{N_c}\ \sum_{K^\prime} \Gamma^{\rm pp} (K|K') {{\bar{\chi}}_0^{\rm pp}}(K') \phi_\alpha (K^\prime) = \lambda_\alpha \phi_\alpha (K) \label{Eq:eigvalcg} 
\end{eqnarray}
with ${{\bar{\chi}}_0^{\rm pp}}(K') = \frac{N_c}{N} \sum_{\tk^{\prime}} G_\uparrow (\K^\prime+\tk^\prime,i\omega_{n'}) \, G_\downarrow (-\K^\prime-\tk^\prime,-i\omega_{n'})$.

\begin{figure}
	[htb] 
	\begin{center}
		\includegraphics*[width=3.2in]{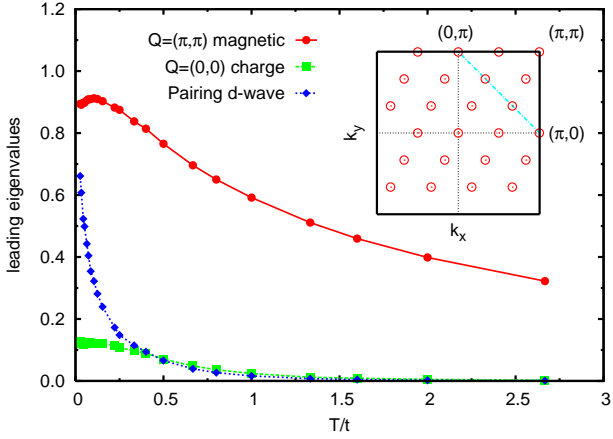} \caption{Leading eigenvalues of the Bethe-Salpeter equation (e.g.\ Eq.~\ref{Eq:eigvalcg}) in various channels for $U/t=4$ and a site occupation $\langle n\rangle=0.85$ calculated on the 24-site cluster shown in the inset. The ${\bf Q}=(\pi, \pi)$, $\omega_m=0$, $S=1$ magnetic eigenvalue saturates at low temperatures. The leading eigenvalue in the singlet ${\bf Q}=(0,0)$, $\omega_m=0$ particle-particle channel has $d_{x^2-y^2}$ symmetry and increases toward 1 at low temperatures \cite{maier:schm}. The largest charge density eigenvalue occurs in the ${\bf Q}=(0, 0)$, $\omega_m=0$ channel and saturates at a small value. (Maier {\it et al.} \cite{maier:pairmech})} \label{fig:eigs} 
	\end{center}
\end{figure}

The temperature dependence of the leading eigenvalues in the pairing, charge density and magnetic channels is plotted in Fig.~\ref{fig:eigs}. Here, $U=4t$ and $\n=0.85$ and the results were obtained with the quantum Monte Carlo dynamic cluster approximation on the 24-site cluster shown in Fig.~\ref{fig:clusters}. An eigenvalue reaches 1 when instabilities to phase transitions occur. As the temperature is lowered, the leading particle-hole eigenvalue occurs in the magnetic channel and has a center of mass momentum $\Q=(\pi,\pi)$ and frequency $\omega_m=0$. This eigenvalue saturates at low temperatures. The leading particle-particle eigenvalue is a spin singlet, and as shown in Fig.~\ref{fig:PhipiTvsK}, its eigenfunction $\Phi_d(\K,\omega_n)$ has $d_{x^2-y^2}$ symmetry. 

\begin{figure}
	[htpb] 
	\begin{center}
		\includegraphics[width=3in]{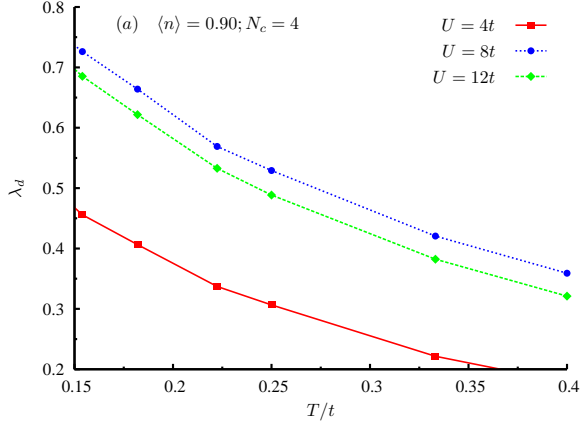}

	\end{center}
	\caption{The $\dw$ eigenvalue $\ld$ versus $T/t$ for $U=4t$, $8t$ and $12t$ and $\n=0.90$. (Maier {\it et al.} \cite{maier:pairint})} \label{fig:lambdad} 
\end{figure}

\begin{figure}
	[htpb] 
	\begin{center}
		\includegraphics[width=3in]{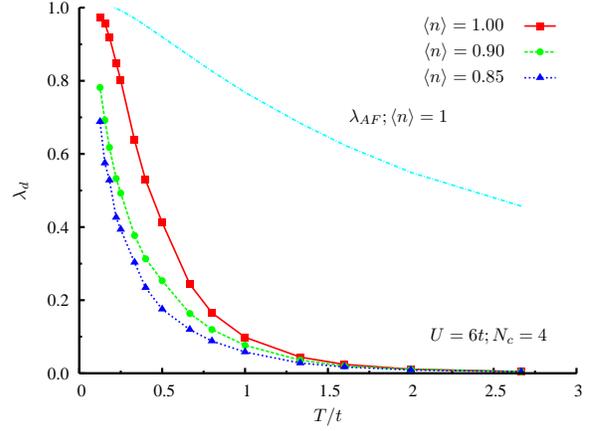} 
	\end{center}
	\caption{The $\dw$ eigenvalue $\ld$ versus $T/t$ for various band fillings $\n$ for $U/t=6$ and $N_c=4$. The dashed line represents the leading eigenvalue $\lambda_{AF}$ in the $\Q=(\pi,\pi)$, $S=1$ particle-hole channel at half-filling. (Maier {\it et al.} \cite{maier:pairint})} \label{fig:lambdadn} 
\end{figure}

\begin{figure}
	[htpb] \centering 
	\includegraphics[width=3in]{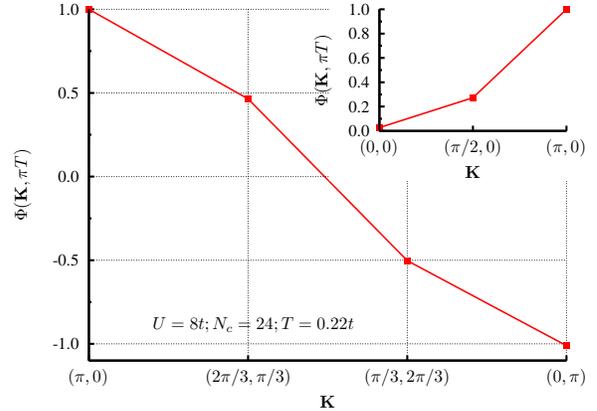} \caption{The $\dw$ eigenvector $\Phi_d(\K, \omega_n)$ at $\omega_n=\pi T$, normalized to its value at $\K = (\pi, 0)$, versus $\K$ for $U/t=8$, band filling $\n=0.9$ and $T/t=0.22$. In the main figure, the $\K$ points move along the dashed line shown in Fig.~\ref{fig:eigs}. The inset shows the behavior of $\Phi_d$ when $\K$ varies along the $k_x$ axis. (Maier {\it et al.} \cite{maier:pairint})} \label{fig:PhipiTvsK} 
\end{figure}

The temperature dependence of the d-wave eigenvalue $\ld$ obtained on a 4-site cluster for $\n=0.9$ with $U/t=4$, 8 and 12 is displayed in Fig.~\ref{fig:lambdad}. The d-wave eigenvalue is largest when $U$ is of order the bandwidth $W=8t$, consistent with the notion that it is important to have strong antiferromagnetic correlations. 
Fig.~\ref{fig:lambdadn} shows $\ld$ for various band fillings $\n$ for $U=6t$ calculated on a 4-site cluster. $\ld$ increases as the system is doped towards half-filling. However, at half-filling the dominant eigenvalue occurs in the particle-hole magnetic channel as indicated by the dashed line, and the groundstate has long-range antiferromagnetic order. Furthermore, for large $U$, a Mott gap opens leaving no holes to pair. 

We now turn to the momentum and frequency dependence of the eigenfunction $\Phi_d(\K,\omega_n)$ that corresponds to the d-wave eigenvalue $\ld$. Fig.~\ref{fig:PhipiTvsK} illustrates the momentum dependence of $\Phi_d(\K,\omega_n)$ for $\omega_n=\pi T$ calculated on a 24-site cluster. Here, $U=8t$ and $T=0.22t$, for which $\lambda_d(T)=0.42$. The values of $\K$ lay along the dashed line shown in the inset of Fig.~\ref{fig:eigs}. Here, the d-wave structure of $\Phi_d(\K,\omega_n)$ is apparent. As shown in the inset, $\Phi_d(\K,\omega_n)$ falls off rapidly for momenta $\K$ moving away from the Fermi surface towards the zone center. 

\begin{figure}
	[htpb] \centering 
	\includegraphics[width=2.8in]{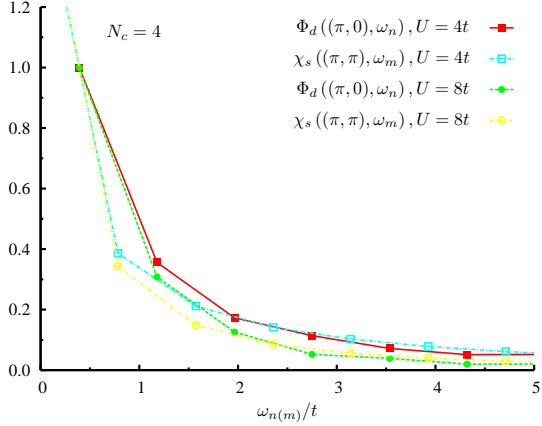} \caption{The Matsubara frequency dependence of $\Phi_d(\K,\omega_n)/\Phi_d(\K,\pi T)$ with $\K=(\pi,0)$ for $U/t=4$ and 8 calculated for $N_c$=4, $T/t=0.125$ and band filling $\n=0.90$. Also shown is the frequency dependence of the normalized spin susceptibility $2\chi_s(\Q,\omega_m)/[\chi_s(\Q,0)+\chi_s(\Q,2\pi T)]$ for $\Q=(\pi, \pi)$. (Maier {\it et al.} \cite{maier:pairint})} \label{fig:PhiChi} 
\end{figure}

The frequency dependence of $\Phi_d(\K,\omega_n)$ at the antinodal point $\K=(\pi,0)$ is plotted in Fig.~\ref{fig:PhiChi} for $U/t=4$ and 8 calculated on a 4-site cluster at $\n=0.9$. Also shown is the spin susceptibility $\chi_s(\Q,\omega_m)$ for $\Q=(\pi,\pi)$ normalized to coincide with $\Phi_d(\K,\omega_n)$ at $\omega_n=\pi T$. One sees that the pairing interaction is retarded and characterized by the same frequency scale as the antiferromagnetic spin susceptibility. As $U$ increases, both $\Phi_d(\K,\omega_n)$ and $\chi_s(\Q,\omega_m)$ fall off more rapidly reflecting the reduction in frequency scale set by the exchange energy. From these results it is clear that the dynamics of the pairing interaction is associated with the spin fluctuation spectrum. 

More can be learned about the mechanism responsible for $d_{x^2-y^2}$ pairing in the doped Hubbard model, by decomposing the pairing interaction $\Gamma^{\rm pp}$ as shown in Fig.~1c. Here, the irreducible particle-particle vertex is given as a combination of a fully irreducible two-fermion vertex $\Lambda_{\rm irr}$ and partially reducible particle-hole exchange contributions \cite{PW89}. For the even frequency, even momentum part of the irreducible particle-particle vertex $\Gamma^{\rm pp}_{\rm even}(K,K')=1/2[\Gamma^{\rm pp}(K,K')+\Gamma^{\rm pp}(K,-K')]$, one obtains 
\begin{eqnarray}
	\Gamma^{pp}_{\rm even} (K|K') &=& \Lambda_{\rm irr} (K|K')\nonumber\\
	&+& \frac{1}{2} \Phi_d(K,K') + \frac{3}{2} \Phi_m (K,K') \label{five} 
\end{eqnarray}

\noindent with $K=(\K,i\omega_n)$. The subscripts $d$ and $m$ denote the charge density $(S=0)$ and magnetic $(S=1)$ particle-hole channels 
\begin{eqnarray}
	\label{six} \Phi_{d/m}(K,K') &=& \frac{1}{2}\left[ \Gamma_{d/m}(K-K';K',-K)\right.\nonumber\\
	&-&\Gamma_{d/m}^{\rm ph}(K-K';K',-K)\nonumber\\
	&+& \Gamma_{d/m}(K+K';-K',-K)\nonumber\\
	&-&\left.\Gamma_{d/m}^{\rm ph}(K+K';-K',-K)\right]\,. 
\end{eqnarray}

\noindent The center of mass and relative wave vectors and frequencies in these channels are labeled by the first, second and third arguments respectively. Using the Monte Carlo results for $\Gpp$, $\Phi_d$ and $\Phi_m$, one can determine the fully irreducible vertex $\Lambda_{\rm irr}$. 
\begin{figure*}
	[htb] 
	\begin{center}
		\includegraphics[width=4.66in]{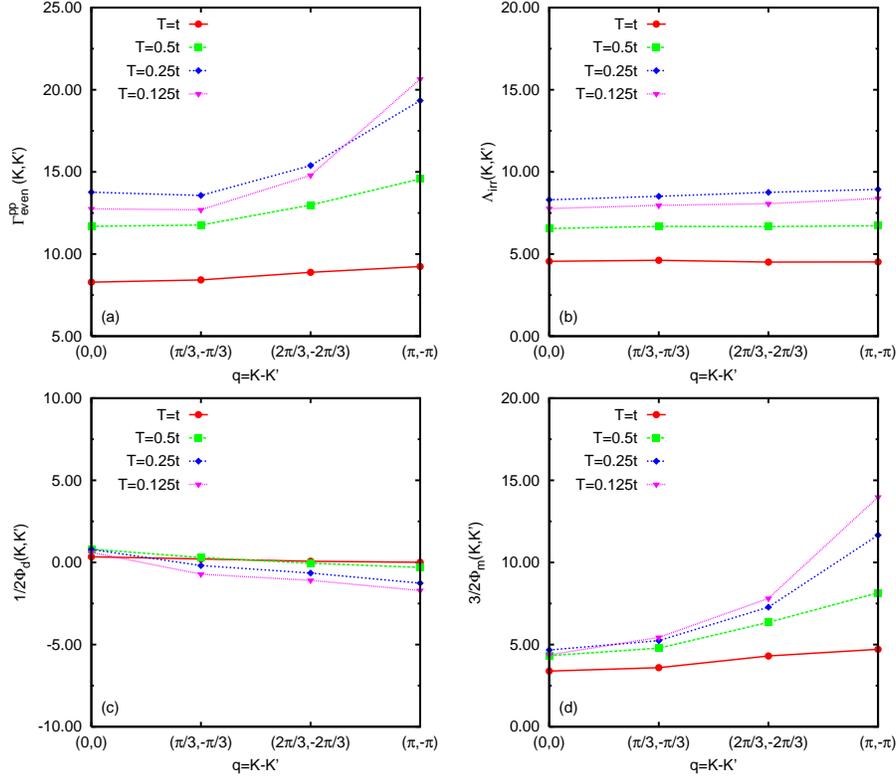} 
	\end{center}
	\caption{(a) The irreducible particle-particle vertex $\Gamma^{\rm pp}(\K|\K')$ versus ${\bf q}=\K-\K^\prime$ for various temperatures with $\omega_n=\omega_{n^\prime}=\pi T$. Here, $\K=(\pi,0)$ and $\K^\prime$ moves along dashed line in the 24-site cluster shown in the inset of Fig.~\ref{fig:eigs}. (b) The ${\bf q}$-dependence of the fully irreducible two-fermion vertex $\Lambda_{\rm irr}$. (c) The ${\bf q}$-dependence of the charge density $(S=0)$ channel $\frac{1}{2}\Phi_d$ for the same set of temperatures. (d) The ${\bf q}$-dependence of the magnetic $(S=1)$ channel $\frac{3}{2} \Phi_m$. (Maier {\it et al} \cite{maier:pairmech})} \label{fig:decomp} 
\end{figure*}

Fig.~\ref{fig:decomp}a shows the irreducible particle-particle vertex $\Gpp(\K|\K')$ versus momentum transfer $\q=\K-\K'$. Here, $\K'$ is set to $(\pi,0)$, $\K$ takes values along the dashed line in Fig.~\ref{fig:eigs} and $\omega_n=\omega_{n'}=\pi T$. When the temperature is lowered, $\Gpp$ increases with momentum transfer $\q$, as one expects for a d-wave pairing interaction. As can be seen from Figs.~\ref{fig:decomp}b-d, the dominant contribution to the pairing interaction $\Gpp$ clearly comes from the magnetic exchange channel, $\Phi_m$, while the charge channel, $\Phi_d$, and the fully irreducible contribution $\Lambda_{\rm irr}$ are rather independent of temperature and momentum transfer.

\section{Conclusions} \label{sec:Conc}

Quantum cluster theories treat correlations within a finite cluster explicitly, while approximating those beyond the cluster at the mean-field level. The 4-site DCA phase diagram for the 2D Hubbard model contains antiferromagnetism, d-wave superconductivity and pseudogap behavior in good agreement with experiments on the cuprates. In large clusters, superconductivity persists and the transition temperature is independent of the cluster size. Results for the pairing interaction $\Gamma^{pp}(\K|\K')$ show that its eigenfunction $\Phi_d(\K,\omega_n)$ of the leading low temperature eigenvalue $\lambda_d$ is an even frequency singlet with $d_{x^2-y^2}$ symmetry. This implies that $\Gpp(\K|\K')$ is short-ranged and increases as the momentum transfer $\q=\K-\K'$ increases. The frequency dependence of $\Phi_d(\K,\omega_n)$ reflects the frequency dependence of the spin susceptibility $\chi_s(\Q,\omega_m)$ for $\Q=(\pi,\pi)$. The leading eigenvalue $\lambda_d$ is largest for $U$ of order the bandwidth and increases towards half-filling, but for large $U$ a Mott-Hubbard gap opens leaving no holes to pair. Finally, an exact decomposition of $\Gpp$ shows that the dominant contribution to the pairing interaction comes from the $S=1$ particle-hole channel.

{\em Acknowledgments} This research was enabled by computational resources of the Center for Computational Sciences at Oak Ridge National Laboratory and conducted at the Center for Nanophase Materials Sciences, which is sponsored at Oak Ridge National Laboratory by the Division of Scientific User Facilities, U.S. Department of Energy.

\end{document}